\documentclass[conference]{IEEEtran}
\IEEEoverridecommandlockouts
\usepackage{cite}
\usepackage{amsmath,amssymb,amsfonts}
\usepackage{algorithmic}
\usepackage{graphicx}
\usepackage{textcomp}
\usepackage{xcolor}
\def\BibTeX{{\rm B\kern-.05em{\sc i\kern-.025em b}\kern-.08em
    T\kern-.1667em\lower.7ex\hbox{E}\kern-.125emX}}
    \makeatletter
\newcommand{\linebreakand}{%
  \end{@IEEEauthorhalign}
  \hfill\mbox{}\par
  \mbox{}\hfill\begin{@IEEEauthorhalign}
}
\makeatother
\begin{document}

\title{A Flexible Architecture for Broadcast Broadband
Convergence in Beyond 5G\\
}


\author{\IEEEauthorblockN{Rashmi Yadav\IEEEauthorrefmark{1},
Rashmi Kamran\IEEEauthorrefmark{2}, Pranav Jha\IEEEauthorrefmark{2} and
Abhay Karandikar\IEEEauthorrefmark{2}\IEEEauthorrefmark{4}}
\IEEEauthorblockA{Department of Electrical Engineering,
Indian Institute of Technology Kanpur, India\IEEEauthorrefmark{1}\\
\IEEEauthorblockA{Department of Electrical Engineering,
Indian Institute of Technology Bombay, India\IEEEauthorrefmark{2}\\
\IEEEauthorblockA{Secretary to the Government of India,
Department of Science \& Technology, New Delhi, India\IEEEauthorrefmark{4}\\
Email: rashmiy@iitk.ac.in\IEEEauthorrefmark{1},
rashmi.kamran@iitb.ac.in\IEEEauthorrefmark{2},
pranavjha@ee.iitb.ac.in\IEEEauthorrefmark{2},
karandi@ee.iitb.ac.in\IEEEauthorrefmark{2}}}}}
\maketitle
\begin{abstract}
\par There has been an exponential increase in the usage of multimedia services in mobile networks in recent years. To address this accelerating data demand, mobile networks are experiencing a subtle transformation in their architecture. One of the changes in this direction is the support of Multicast/Broadcast Service (MBS) in the Third Generation Partnership Project (3GPP) Fifth Generation (5G) network. The MBS has been introduced to enhance resource utilization and user experience in 3GPP 5G networks. However, there are certain limitations in the 3GPP 5G MBS architecture, such as the selection of the delivery method (unicast or broadcast) by the core network (may result in sub-optimal radio resource utilization) and no provision for converging non-3GPP broadcast technologies (like digital terrestrial television) with cellular (3GPP 5G) broadband. In this context, we propose a new architecture for the convergence of cellular broadband and non-3GPP broadcast networks. A novelty of the architecture is that it treats signalling exchange with User Equipment (UE) as data (service) which results in improved scalability of mobile networks. The architecture supports enhanced flexibility in choosing a delivery method (3GPP 5G unicast, 3GPP 5G broadcast, or non-3GPP broadcast) for user data. We evaluate the performance of the proposed architecture using process algebra-based simulations, demonstrating a significant reduction in the number of signalling messages exchanged between the UE and the network for MBS session establishment as compared to the 3GPP 5G network. 

\end{abstract}


\section{Introduction}
\IEEEPARstart{I}n recent years, there has been a remarkable rise in multimedia content utilization over mobile networks. As highlighted in the Ericsson mobility report \cite{b1}, video content, especially in social media and video-on-demand services, comprises the largest and fastest-growing segment of mobile data traffic globally, which accounts for approximately 70\% of traffic share in 2022. Highlights from the same report show an annual growth of about 30\% by the end of 2028, further increasing the global mobile data traffic's video share to 80\%. Moreover, Qualcomm's broadcast report \cite{b2} predicts an enormous rise in live streaming content on social media, with approximately 800 million users expected to participate in daily live streams. These statistics reflect the significant impact and importance of multimedia services in the present-day mobile network.

The deployment of Fifth Generation (5G) mobile networks brings the possibility to increase the usage of Multicast/Broadcast Service (MBS). MBS is a crucial use case to address the increasing data demands within the framework of 5G technology. Apropos to this, Release 17 of the Third Generation Partnership Project (3GPP) 5G standards has introduced the support for MBS to enhance the 3GPP 5G architecture. Nevertheless, there are some architectural limitations of the 3GPP 5G MBS support. These include the selection of delivery methods (unicast or broadcast) by the Core Network (CN) that might result in sub-optimal utilization of resources, limited handling for user mobility for MBS, and no provision for the convergence of Non-3GPP Broadcast Networks (N3BNs) within the 3GPP 5G (such as Digital Terrestrial Television (DTT)). 

\par This section provides the literature survey related to MBS architecture and mechanisms. The work in \cite{b11} presents various architectural concepts and mechanisms to optimize network loading and traffic patterns for MBS delivery. The paper \cite{b24} presents a comprehensive review on the convergence of broadcast and broadband in the 5G network. The authors in \cite{b23} proposed an enhanced Next Generation Radio Access Network (NG-RAN) architecture with architectural and functional enhancements to provide the efficient delivery of terrestrial broadcast services. The work in \cite{b15} explores a mixed transmission mode that utilizes shared multicast, broadcast, and unicast resources over the same physical channel. In \cite{b17}, authors review the upcoming 3GPP 5G standards, discuss limitations of 3GPP 5G MBS architecture and present state-of-the-art standardization initiatives towards integrating N3BNs with the 5G. Furthermore, the latest release of the 3GPP standard (Release 18) does not include the support for integration of the N3BN and the 3GPP broadcast network.
\par To the best of our knowledge, the prior art lacks comprehensive architectural solutions related to the convergence of cellular broadband (3GPP 5G) and N3BNs and scalability enhancements in the context of MBS delivery in mobile networks. We propose a flexible architecture for broadcast broadband convergence where we treat UE signalling as data or service. Therefore, we call the proposed architecture a Signalling Service-Based Architecture (SSBA) for broadcast broadband convergence. This work is based on prior research conducted in \cite{b21}, focusing on the convergence of broadcast and broadband as the authors did not delve into this specific aspect. The SSBA presents a scalable network architecture making it a promising solution for the Beyond 5G (B5G) landscape through enhanced resource utilization in a converged network. SSBA provides the flexibility to choose between broadcast and broadband delivery methods based on resource availability. We are exploring LTE and 5G broadcast (FeMBMS), but the proposed architecture can also incorporate other broadcast schemes, such as the Advanced Television Systems Committee (ATSC). An illustration of the convergence of N3BN in the proposed SSBA is provided in Section \ref{dtt}. In Section \ref{perf}, we evaluate the performance of the proposed SSBA using the Eclipse plug-in \cite{b20}, a tool for modelling distributed systems with the help of Performance Evaluation Process Algebra (PEPA) \cite{b19}, a modelling language. 


\par The rest of the paper is as follows: Section \ref{arch} also presents the architectural details of the proposed signalling service-based architecture. Section \ref{model} presents the system model. We conclude in Section \ref{conc} along with future directions.
\section{Proposed Signalling Service-Based Architecture}
\label{arch}
\begin{figure}[htbp]
\hspace{0.1cm}
 \centering
\includegraphics[scale=0.6]{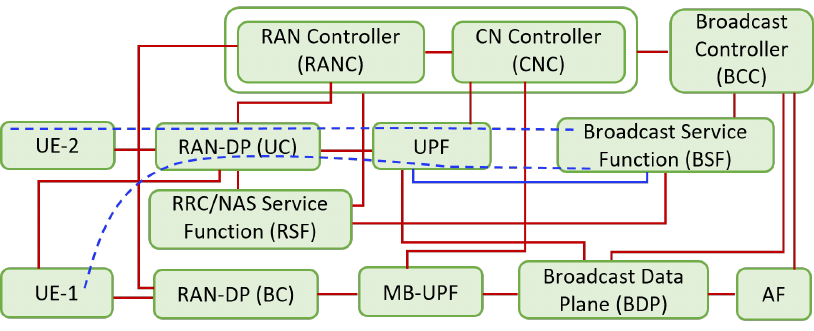}
\caption{Proposed signalling service-based mobile network architecture.}
\label{mbs_pro}
\vspace{-0.1cm}
\end{figure}

In this section, we present an overview of the proposed SSBA, as shown in Fig.\,\ref{mbs_pro}. Before delivering the details of the proposed SSBA, let us first understand the basic principles of an SSBA (a detailed explanation is available in \cite{b21}). In SSBA, a Service Function (SF) handles signalling exchange with the UE through the user plane for network services (i.e. RRC/NAC service, broadcast service). The SF then communicates with the network control plane to establish the data path through the network data plane.
 
\par In the context of Fig.\ref{mbs_pro}, the preceding explanation can be linked as follows: the Broadcast Service Function (BSF) functions as an independent SF responsible for exchanging multicast/broadcast-related signalling messages with UEs. These include membership requests, content delivery requests, and other associated operations. The BSF interfaces with the control plane (Broadcast Controller (BCC)) to initiate data path establishment. Further, the BCC communicates with the RAN/CN controller to configure the data path for efficient content delivery. However, the decision for delivery methods (unicast or broadcast) is governed by BCC and switching between delivery methods is handled by the Broadcast Data Plane (BDP).
\par The information flow in the proposed SSBA is as follows: UE-1 is associated with the broadcast RAN (RAN-DP (BC)), whereas UE-2 is connected to the unicast (UC) RAN (RAN-DP (UC)). Notably, the BSF facilitates UE (both UE-1 and UE-2) interaction via the DP. Therefore, the signalling path for UC/BC transmission involves the following: UE-2/UE-1 - RAN-DP (UC) - User Plane Function (UPF) - BSF as shown in Fig.\,\ref{mbs_pro} with the blue dotted line. Furthermore, the RAN controller controls the RAN data planes (RAN-DP (UC) and RAN-DP (BC)), while the CN controller supervises the CN data planes (UPF and Multicast/Broadcast UPF (MB-UPF)).

\subsection{MBS Session Establishment Call flow for the Proposed SSBA}
\begin{figure*}[htbp]
	\centering
	\hspace{-0.6cm}
    \centering
	\includegraphics[scale=0.61]{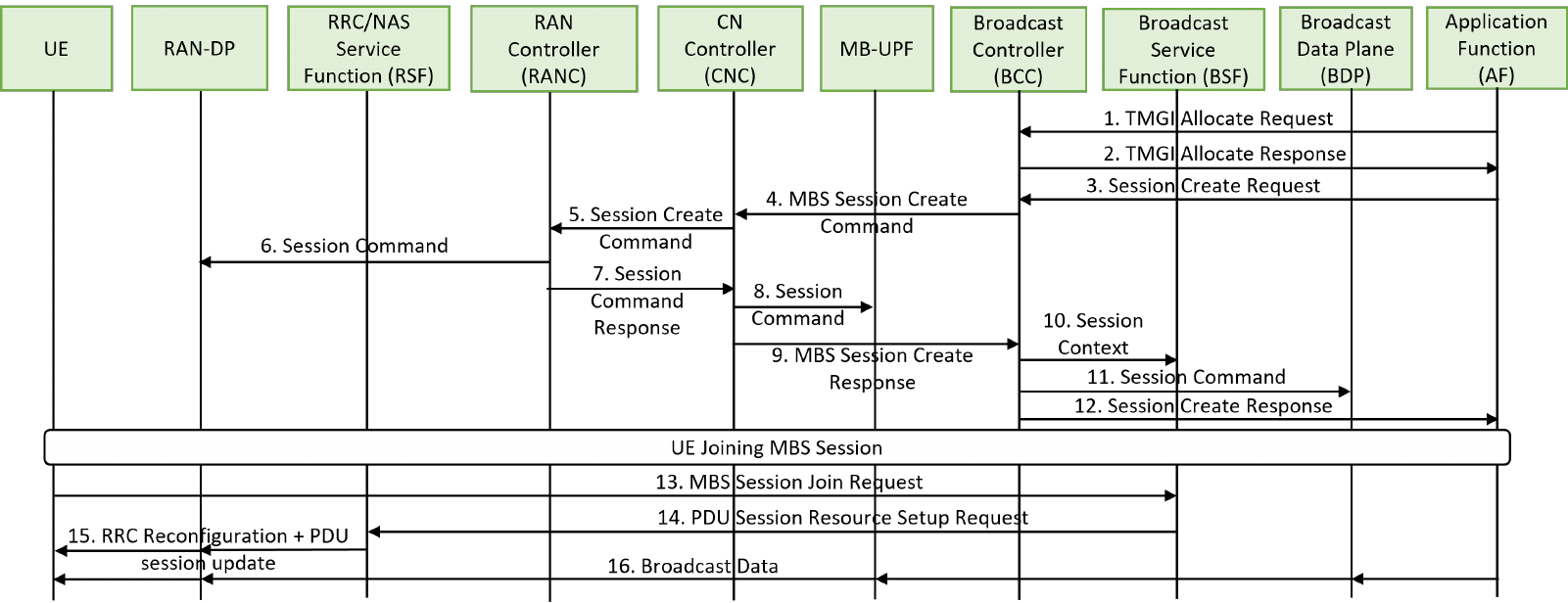}
	\caption{MBS session establishment call flow in the proposed SSBA.}
	\label{call_flow_pro}
	\vspace{-0.3cm}
\end{figure*}
Fig.\,\ref{call_flow_pro} illustrates the MBS session establishment call flow for the proposed SSBA. Conversely, the call flow for MBS session establishment in the 3GPP 5G architecture is available in \cite{b3} (Section 7.1.1.2 and 7.2.1.3). This section provides a concise comparison between the call flows of the proposed and 3GPP 5G MBS session establishment, highlighting the significant reduction in the number of signalling messages achieved in the proposed SSBA. 
\par A Temporary Mobile Group Identity (TMGI) allocation process (messages 1-2) is carried out between Application Function (AF) and BCC as shown in Fig.\,\ref{call_flow_pro}. The session create request (message 3) is sent to BCC by AF. In response, BCC forwards this request to the CN controller. CN controller confirms the resource availability from the RAN controller (messages 5 and 7) and the RAN controller sends the session command (message 6) to RAN-DP to set up the session on the RAN side. On receiving confirmation from the RAN controller, the CN controller sends the session command to MB-UPF (message 8) to set up the session on the core network side as well. Messages 13-15 are related to UE joining the MBS session. Here, UE initiates the MBS session join request to BSF, which is then forwarded to the RRC/NAS SF and eventually processed by this SF to send the Radio Resource Control (RRC) reconfiguration message in response, to UE via RAN-DP.
\par The controllers in the proposed SSBA do not call for response messages from the user plane, as they have global information about the user plane resources. Therefore, messages 6, 8 and 11 (as shown in Fig.\,\ref{call_flow_pro}) are sent as commands without requiring corresponding response messages. As a result, the modified call flow significantly reduces the total number of signalling messages. Similarly, message 14 (Fig.\,\ref{call_flow_pro}) also does not require a response message. Therefore, the procedure for MBS session establishment call flow is appreciably simplified, enhancing the overall modularity of the network.

\section{System Model}
\label{model}
In this section, we describe the system modelling of the proposed SSBA, using the PEPA \cite{b19}, a high-level language, for modelling distributed systems. The PEPA modelling facilitates system performance evaluation using the Eclipse plug-in \cite{b20}, a tool for performance analysis. In Table\,\ref{tab:table1}, we present the MBS session establishment call flow (Fig.\,\ref{call_flow_pro}) modelling.

\par To explain the behaviour of various Network Functions (NFs) involved in the MBS session establishment call flow, we model each NF as a PEPA component. This approach allows us to represent the tasks executed by each PEPA component/NF through different states. Each state is denoted by the NF's name followed by a number corresponding to its state ($NF_1$). For instance, the NF \textit{UE} possesses two states, $Ue_1$ and $Ue_2$, representing different states of the UE to perform various tasks. As shown in the table, similar representations of states are provided for each NF, such as RAN-DP, SF, RAN controller, CN controller, MB-UPF, BCC, BSF, BDP, and AF.

\par Furthermore, we model each task of the NF as an action type, indicated in lowercase letters. To provide further details, subscripts are added to the action types ($actiontype_{detail}$). For instance, the action type $tmgi_{req}$ signifies the TMGI allocation request, where $tmgi$ is the action type, and $req$ corresponds to its specific detail as a $request$. Each action type is also associated with a rate value, denoted as $r$. These rates represent the expected duration of a particular action type in the PEPA component and values are considered as given in \cite{b5}, \cite{b6}, and \cite{b7}.
\par By using PEPA modelling, we can analyze the performance of the proposed SSBA for MBS session establishment call flow. To understand the system modelling, consider an example of SF as an NF to model it as a PEPA component as shown in Table\,\ref{tab:table1}. This NF (SF) is represented with two states: $Sf_1$ and $Sf_2$, each performing specific tasks (as shown in Fig.\,\ref{call_flow_pro}). In the first state, $Sf_1$, the action type $setup_{req}$ signifies the ``PDU session resource setup request'' as a task performed during this state.

\par Moving to the second state, $Sf_2$, two actions are being performed within this state. The first action, denoted as $get_{sfp}$, represents the SF NF's attempt to access the SF processor (SFP) to process the received request. The second action, denoted as $reconfig$, represents the specific action of RRC reconfiguration involving the PDU session modification command. For the processing of requests, each NF is associated with a corresponding processor. For this purpose, we define processors using a two-state model as defined in \cite{b8}, \cite{b9}. However, the following are the associated processors considered in the example: UE processor (\textit{UEP}), RAN-DP processor (\textit{RANDPP}), \textit{SFP}, RAN controller processor (\textit{RANP}), CN controller processor (\textit{CNP}), MB-UPF processor (\textit{MB-UPFP}), BCC processor (\textit{BCCP}), BSF processor (\textit{BSFP}), BDP processor (\textit{BDPP}) and AF processor (\textit{AFP}). For instance, the AFP is defined with two states. The first state, $Afp_1$, represents the action of getting access to the AF processor ($get_{afp}$), while the second state is dedicated to performing actions associated with the processor ($tmgireq$ and $sessionreq$). Similarly, other processors (as shown in Table\,\ref{tab:table1}) are defined with their corresponding NFs.

\par The system equation (as shown in Table\,\ref{tab:table1}) describes the interactions between the NFs in the proposed SSBA. These interactions are performed between various NFs as $NF_1$ $_{V}^{\bowtie}$ $NF_2$, where $V$ = $<$$action_1$, $action_2$$>$ is a set of actions performed between $NF_1$ and $NF_2$. For instance, consider the example $Randp_1$[N] $_{V_2}^{\bowtie}$ $Ran_1$[N] which signifies that NF $Ran_1$ interacts with $Randp_1$ NF through the set of actions contained in variable $V_2$. Given $V_2$ = $<$$sessionrandp$$>$, implies the interaction between $Ran_1$ and $Randp_1$ to perform the action type $sessionrandp$ (signifies that the RAN controller commands the RAN-DP to set up a session). In a similar way, the interactions between other NFs are modelled and presented in Table\,\ref{tab:table1}. 

\par Moreover, several other variables are used in the system equation that require further discussion. The variable $n$ represents the number of UEs. $N_{nf}$ denotes the number of specific NFs, such as $N_{randp}$, $N_{sf}$, $N_{ran}$, $N_{cn}$, $N_{mbupf}$, $N_{bcc}$, $N_{bsf}$, $N_{bdp}$, and $N_{af}$, representing the number of RAN-DP, SF, RAN controller, CN controller, MB-UPF, BCC, BSF, BDP, and AF NFs, respectively. Furthermore, each processor can handle a set of concurrent threads, denoted as $N_t$, while the number of processors allocated to each NF is represented as $N_{nfp}$. Consequently, $N$ = $N_{nf}$.$N_{nfp}$.$N_t$ presents the total number of threads for a specific NF, and $N_p$ = $N_{nf}$.$N_{nfp}$ denotes the total number of processors allocated to a particular NF type.

\par However, we followed a similar modelling procedure and simulations were then performed for both the 3GPP 5G and the proposed SSBA's MBS session establishment call flow, demonstrating a comparative analysis of their performance in terms of the reduced number of signalling messages, enhanced modularity and improved scalability.

\begin{table}
\caption{System modelling for MBS session establishment call flow\label{tab:table1}}
\centering
\vspace{0.1cm}
\fontsize{7.5pt}{7.5pt}\selectfont
\begin{tabular}{|p{0.18\columnwidth}|p{0.7\columnwidth}|}
\hline
\textbf{PEPA Modelling of NFs} & \textbf{\centering Code Description}\\
\hline
UE NF & \textit{$Ue_1$ $_{=}^{def}$ ($get_{uep}$, $r_p$).($mbsjoinreq$, $r_{iat}$).$Ue_2$} \\
& \textit{$Ue_2$ $_{=}^{def}$ ($reconfig_1$, $r_v$).$Ue_1$} \\
\hline
RAN-DP & \textit{$Randp_1$ $_{=}^{def}$ ($sessionrandp$,$r_v$).$Randp_2$} \\
NF & \textit{$Randp2$ $_{=}^{def}$ ($get_{randpp}$,$r_p$).($prepare$,$r_v$).$Randp_1$} \\
\hline
RSF  & \textit{$Rsf_1$ $_{=}^{def}$ ($setup_{req}$,$r_v$).$Rsf_2$} \\
NF & \textit{$Rsf_2$ $_{=}^{def}$ ($get_{rsfp}$,$r_p$).($reconfig$,$r_v$).$Rsf_1$} \\
\hline
RAN & \textit{$Ran_1$ $_{=}^{def}$ ($sessioncreate$,$r_v$).$Ran_2$} \\
controller NF & \textit{$Ran2$ $_{=}^{def}$ ($get_{ranp}$,$r_p$).($sessionrandp$,v)} \\
& .($sessionres$,v).$Ran_1$ \\
\hline
CN  & \textit{$Cn_1$ $_{=}^{def}$ ($mbsessioncom$,$r_v$).$Cn_2$} \\
controller NF & \textit{$Cnc_2$ $_{=}^{def}$ ($get_{cnp}$,$r_p$).($sessioncreate$,$r_v$).$Cn_3$} \\
& \textit{$Cn_3$ $_{=}^{def}$ ($sessionres$,$r_v$).$Cnc_4$} \\
& \textit{$Cn_4$ $_{=}^{def}$ ($get_{cnp}$,$r_p$).($sessionupf$,$r_v$)} \\
& .($mbsessionres$,$r_v$).$Cn_1$ \\
\hline
MB-UPF NF & \textit{$Mbupf_1$ $_{=}^{def}$ ($sessionupf$,$r_v$).$Mbupf_2$} \\
& \textit{$Mbupf_2$ $_{=}^{def}$ ($get_{mbupfp}$,$r_p$)}.($prepare$,$r_v$).$Mbupf_1$ \\
\hline
BCC NF & \textit{$Bcc_1$ $_{=}^{def}$ ($tmgireq$,$r_v$).$Bcc_2$} \\
& \textit{$Bcc_2$ $_{=}^{def}$ ($get_{bccp}$,$r_p$).($tmgires$,$r_v$).$Bcc_3$} \\
& \textit{$Bcc_3$ $_{=}^{def}$ ($sessionreq$,$r_v$).$Bcc_4$} \\
& \textit{$Bcc_4$ $_{=}^{def}$ ($get_{bccp}$,$r_p$).($mbsessioncomm$,$r_v$)}. \\
& \textit{$Bcc_4$ $_{=}^{def}$ ($mbsessionres$,$r_v$).$Bcc_5$} \\
& \textit{$Bcc_5$ $_{=}^{def}$ ($get_{bccp}$,$r_p$).($sessioncontext$,$r_v$)} \\
& .($sessionbdp$,$r_v$).($sessionres$,$r_v$).$Bcc_1$ \\
\hline
BSF NF & \textit{$Bsf_1$ $_{=}^{def}$ ($sessioncontext$,$r_v$).($mbsjoinreq$,$r_v$).$Bsf_2$} \\
& \textit{$Bsf_2$ $_{=}^{def}$ ($get_{bsfp}$,$r_p$).($setupreq$,$r_v$).$Bsf_1$} \\
\hline
BDP NF & \textit{$Bdp_1$ $_{=}^{def}$ ($sessionbdp$,$r_v$).$Bdp_2$} \\
& \textit{$Bdp_2$ $_{=}^{def}$ ($get_{bdpp}$,$r_p$).($prepare$,$r_v$).$Bdp_1$} \\
\hline
AF NF & \textit{$Af_1$ $_{=}^{def}$ ($get_{afp}$,$r_p$).($tmgireq$,$r_v$).$Af_2$} \\
& \textit{$Af_2$ $_{=}^{def}$ ($tmgires$,$r_v$).$Af_3$} \\
& \textit{$Af_3$ $_{=}^{def}$ ($get_{afp}$,$r_p$).($sessionreq$,$r_v$).$Af_4$} \\
& \textit{$Af_4$ $_{=}^{def}$ ($sessionres$,$r_v$).$Af_1$} \\
\hline
UE Processor & \textit{$Uep_1$ $_{=}^{def}$ ($get_{uep}$, $r_p$).$Uep_2$} \\
& \textit{$Uep_2$ $_{=}^{def}$ ($mbsjoinreq$,$r_{iat}$).$Uep_1$} \\
\hline
RAN-DP & \textit{$Randpp_1$ $_{=}^{def}$ ($get_{randpp}$, $r_p$).$Randpp_2$} \\
Processor & \textit{$Randpp_2$ $_{=}^{def}$ ($prepare$,$r_v$).$Randpp_1$} \\
\hline
RSF & \textit{$Rsfp_1$ $_{=}^{def}$ ($get_{rsfp}$, $r_p$).$Rsfp_2$} \\
Processor & \textit{$Rsfp_2$ $_{=}^{def}$ ($reconfig$,$r_v$).$Rsfp_1$} \\
\hline
RAN & \textit{$Ranp_1$ $_{=}^{def}$ ($get_{rancp}$, $r_p$).$Ranp_2$} \\
Processor & \textit{$Ranp_2$ $_{=}^{def}$ ($sessionrandp$,$r_v$).$Ranp_1$} \\
& +($sessionres$,$r_v$).$Ranp_1$ \\
\hline
CN Processor  & \textit{$Cnp_1$ $_{=}^{def}$ ($get_{cnp}$,$r_p$).$Cnp_2$} \\
& \textit{$Cnp_2$ $_{=}^{def}$ ($sessioncreate$,$r_v$).$Cnp_1$} \\
& +($sessionupf$,$r_v$).$Cnp_1$+($mbsessionres$,$r_v$).$Cnp_1$ \\
\hline
MB-UPF & \textit{$Mbupfp_1$ $_{=}^{def}$ ($get_{mbupfp}$,$r_p$).$Mbupfp_2$} \\
Processor  & \textit{$Mbupfp_2$ $_{=}^{def}$ ($prepare$,$r_v$).$Mbupfp_1$} \\
\hline
BCC & \textit{$Bccp_1$ $_{=}^{def}$ ($get_{bccp}$,$r_p$).$Bccp_2$} \\
Processor & \textit{$Bccp_2$ $_{=}^{def}$ ($tmgires$,$r_v$).$Bccp_1$}+($mbsessioncom$,$r_v$) \\
& .$Bccp_1$+($sessioncontext$,$r_v$).$Bccp_1$+($sessionres$,$r_v$) \\
& .$Bccp_1$+($sessionbdp$,$r_v$).$Bccp_1$ \\
\hline
BSF & \textit{$Bsfp_1$ $_{=}^{def}$ ($get_{bsfp}$,$r_p$).$Bsfp_2$} \\
Processor & \textit{$Bsfp_2$ $_{=}^{def}$ ($setupreq$,$r_v$).$Bsfp_1$} \\
\hline
BDP & \textit{$Bdpp_1$ $_{=}^{def}$ ($get_{bdpp}$,$r_p$).$Bdpp_2$} \\
Processor & \textit{$Bdpp_2$ $_{=}^{def}$ ($prepare$,$r_v$).$Bdpp_1$} \\
\hline
AF Processor  & \textit{$Afp_1$ $_{=}^{def}$ ($get_{afp}$,$r_p$).$Afp_2$} \\
& \textit{$Afp_2$ $_{=}^{def}$ ($tmgireq$,$r_v$).$Afp_1$+($sessionreq$,$r_v$).$Afp_1$} \\
\hline
System & $Ue_1$[n]$_{\phi}^{\bowtie}$$Randp_1$[N]$_{V_1}^{\bowtie}$$Rsf_1$[N]$_{V_2}^{\bowtie}$$Ran_1$[N]$_{V_3}^{\bowtie}$$Cn_1$[N])\\
Equation & $_{V_4}^{\bowtie}$$Mbupf_1$[N]$_{V_5}^{\bowtie}$$Bcc_1$[N]$_{V_6}^{\bowtie}$$Bsf_2$[N]$_{V_7}^{\bowtie}$$Bdp_1$[N] \\
& $_{V_8}^{\bowtie}$$Af_2$[N]$_{V_9}^{\bowtie}$ $Uep_1$[n] $_{\phi}^{\bowtie}$ $Randpp_1$[$N_p$]$_{\phi}^{\bowtie}$ $Rsfp_1$[$N_p$]  \\ 
& $_{\phi}^{\bowtie}$ $Ranp_1$[$N_p$] $_{\phi}^{\bowtie}$ $Cnp_2$[$N_p$]$_{\phi}^{\bowtie}$$Mbupfp_1$[$N_p$]) \\
& $_{\phi}^{\bowtie}$$Bccp_1$[$N_p$])$_{\phi}^{\bowtie}$$Bsfp_2$[$N_p$])$_{\phi}^{\bowtie}$$Bdp_1$[$N_p$]$_{\phi}^{\bowtie}$$Afp_2$[$N_p$] \\ 
\hline
Variables & $V_1$ = $<$$reconfig$$>$ \\ 
& $V_2$ = $<$$sessionrandp$$>$ \\ 
& $V_3$ = $<$$sessioncreate$,$sessionres$$>$ \\ 
& $V_4$ = $<$$sessionupf$$>$\\
& $V_5$ = $<$$mbsessioncom$,$mbsessionres$$>$ \\ 
& $V_6$ = $<$$mbsjoinreq$,$setupreq$$>$ \\
& $V_7$ = $<$$sessionbdp$$>$ \\
& $V_8$ = $<$$tmgireq$,$tmgires$,$sessionreq$,$sessionres$$>$ \\ 
& $V_9$ = $<$$get_{uep}$, $get_{randpp}$, $get_{rsfp}$, $get_{ranp}$, $get_{cnp}$,\\
& $get_{mbupfp}$, $get_{bccp}$, $get_{bsfp}$, $get_{bdpp}$, $get_{affp}$$>$ \\
& \textit{$\phi$} = $<>$ \\
\hline
\end{tabular}
\vspace{-0.3cm}
\end{table}

\section{Performance Evaluation}
\label{perf}
In this section, we present the performance evaluation of both the proposed SSBA and the 3GPP 5G architecture using the Eclipse plug-in tool \cite{b20}. The evaluation is based on several parameters, such as the number of MBS sessions established per unit time, average response time (ART), and processor utilization. These parameters are significant in evaluating the network's scalability, one of the key aspects we consider. The MBS session establishment rate measures the rate at which MBS sessions are established with respect to specific actions, such as $reconfig$ representing RRC reconfiguration (PDU session modification command). This specific action represents the completion of the MBS session establishment call flow. ART evaluates the average waiting time for UE's MBS session establishment process. Processor utilization evaluates the NF's processor capacity utilization during the entire process.

However, with separate controllers, user plane functions, broadcast service functions and other NFs, the proposed SSBA can be viewed as a distributed system similar to the 3GPP 5G architecture, we propose to use the scalability parameter of a distributed system to evaluate and perform the comparison of their scalability. The scalability (\textit{S}) for a distributed system, based on productivity as defined in \cite{b10}, is given by the ratio of the productivity of the system at two different configurations with different scales $b_1$ and $b_2$ \cite{b6}. In this context, the configurations ($b_1$ and $b_2$) refer to the different numbers of NFs used in the network. For example, $b_1$ = (1,1,1,1,1,1,1,1,1) and $b_2$ = (3,3,3,3,3,3,3,3,3). Configuration $b_1$ represents the basic configuration with a single NF assigned to RAN-DP, SF, RAN controller, CN controller, BCC, MB-UPF, BSF, BDP and AF in the proposed SSBA (i.e.,$N_{randp}$,$N_{sf}$,$N_{ran}$,$N_{cn}$,$N_{mbupf}$,$N_{bcc}$,$N_{bsf}$,$N_{bdp}$,$N_{af}$) = (1,1,1,1,1,1,1,1,1). On the other hand, configuration $b_2$ represents a scaled system with three NFs assigned as follows: ($N_{randp}$,$N_{sf}$,$N_{ranc}$,$N_{cnc}$,$N_{mbupf}$,$N_{bcc}$,$N_{bsf}$,$N_{bdp}$,$N_{af}$) = (3,3,3,3,3,3,3,3,3). Please note that we provide an equal number of total processors in each case (3GPP 5G and the proposed SSBA). The mathematical expression for scalability can be referred from \cite{b22}. By evaluating scalability, we can compare the performance of the proposed SSBA and the 3GPP 5G architecture under different scaling configurations which demonstrate the efficient handling of the proposed SSBA for the increased number of users.

\begin{figure}[h!]
	\centering
    \vspace{-0.5cm}
	\includegraphics[width=0.95\columnwidth]{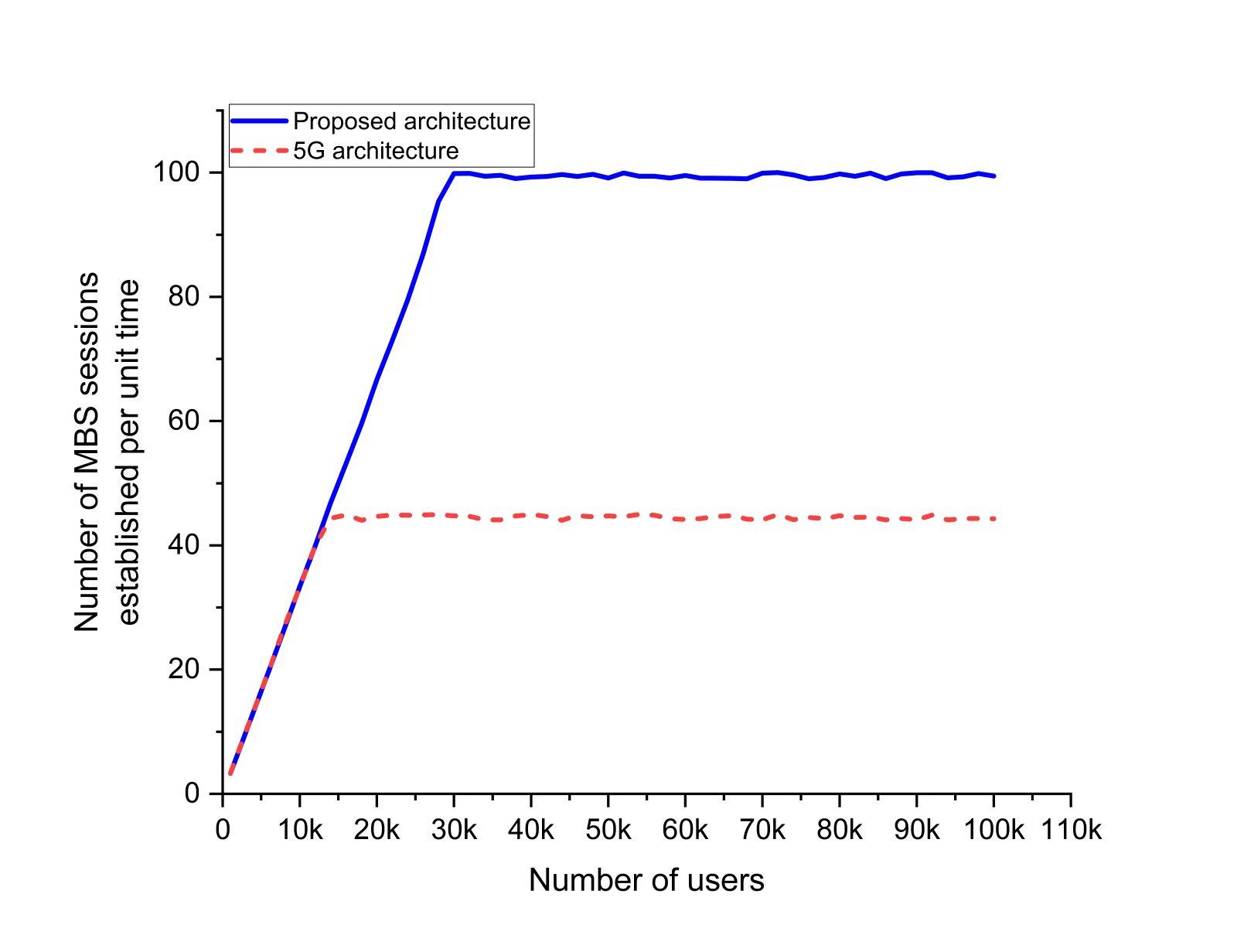}
	\vspace*{-0.3cm}
	\caption{Number of MBS sessions established per unit time for the basic configuration.}
	\label{serv_throu_1}
	\vspace{-0.3cm}
\end{figure}

\begin{figure}[h!]
	\centering
    \vspace{-0.5cm}
	\includegraphics[width=0.95\columnwidth]{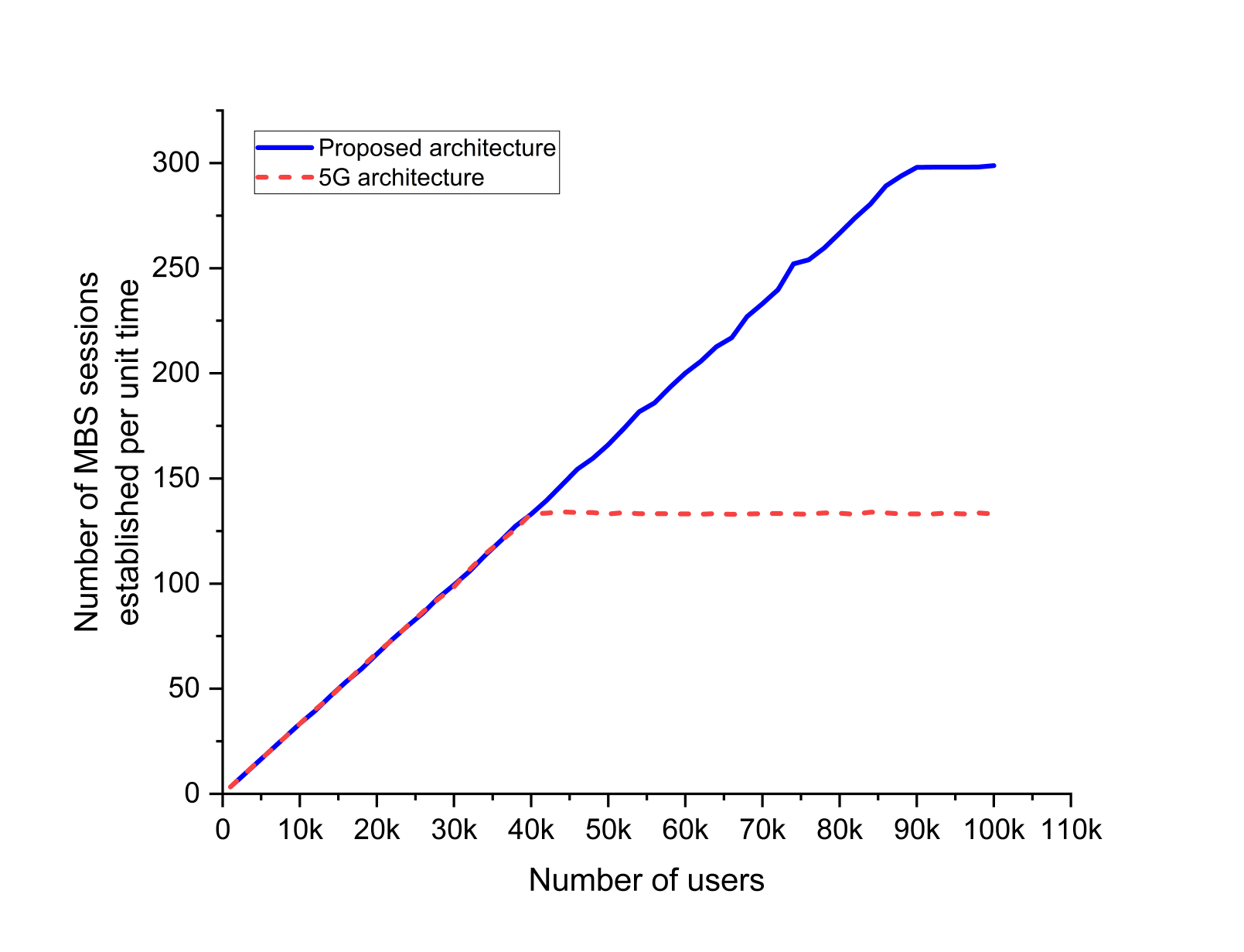}
	\vspace*{-0.3cm}
	\caption{Number of MBS sessions established per unit time for the scaled configuration.}
	\label{serv_throu_3}
	\vspace{-0.1cm}
\end{figure}

Fig. \ref{serv_throu_1} and \ref{serv_throu_3} illustrate the number of MBS sessions established per unit time for both architectures under two different configurations, denoted as $b_1$ and $b_2$. 
It is observed that the proposed SSBA achieves a higher saturation point than the 3GPP 5G architecture. In the basic configuration ($b_1$), the 3GPP 5G architecture saturates at 14,000 users, while the proposed SSBA saturates at 30,000 users. Similarly, in the scaled configuration ($b_2$), the 3GPP 5G architecture saturates at 42,000 users, while the proposed SSBA saturates at 90,000 users. The saturation point indicates the maximum number of UEs served by the network before it becomes overloaded.

\begin{figure}[h!]
	\centering
	\includegraphics[width=0.95\columnwidth]{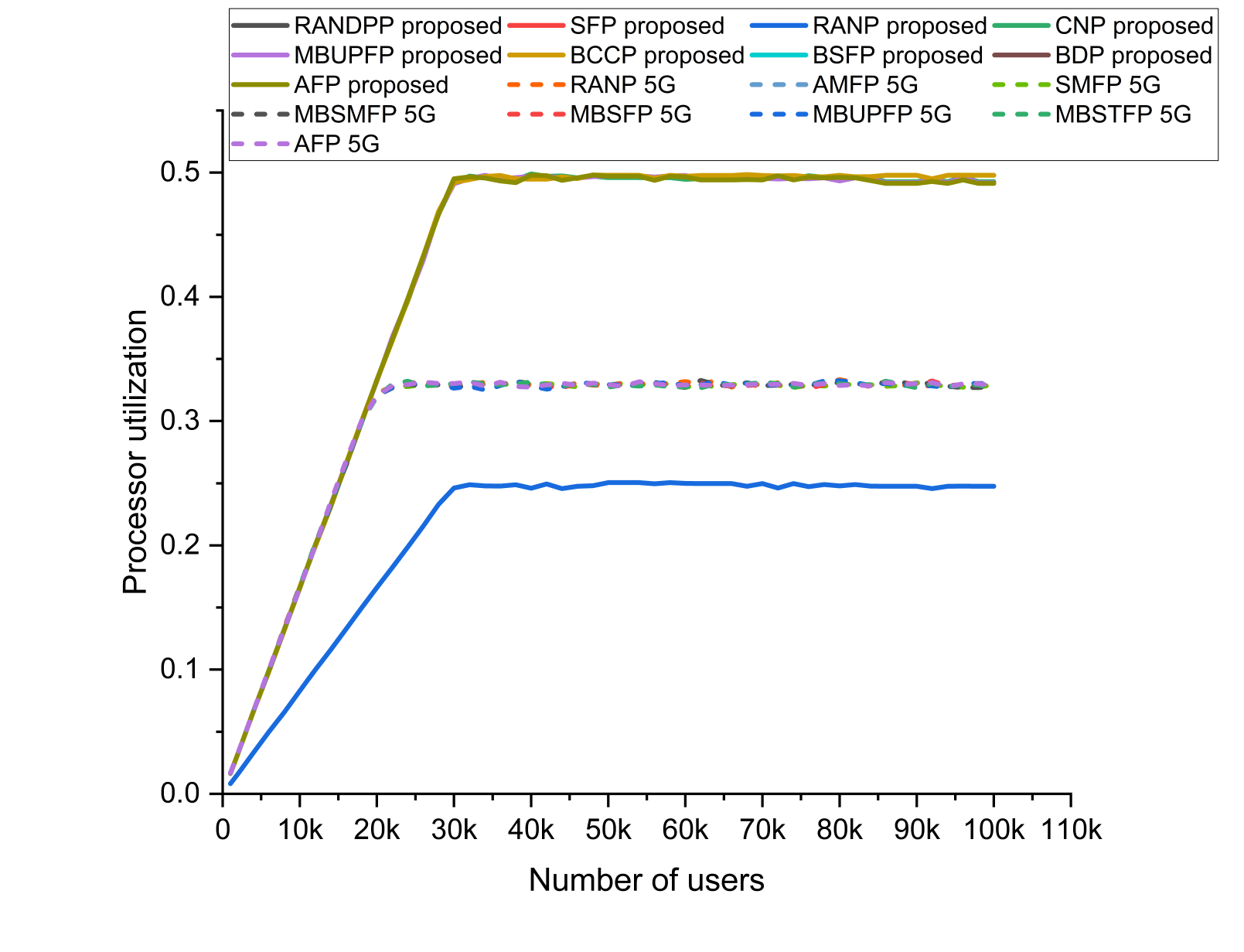}
	\vspace*{-0.3cm}
	\caption{Processor utilization of MBS session establishment for the basic configuration.}
	\label{serv_util_1}
	\vspace{-0.3cm}
\end{figure}

\begin{figure}[h!]
	\centering
    \vspace{-0.1cm}
	\includegraphics[width=0.95\columnwidth]{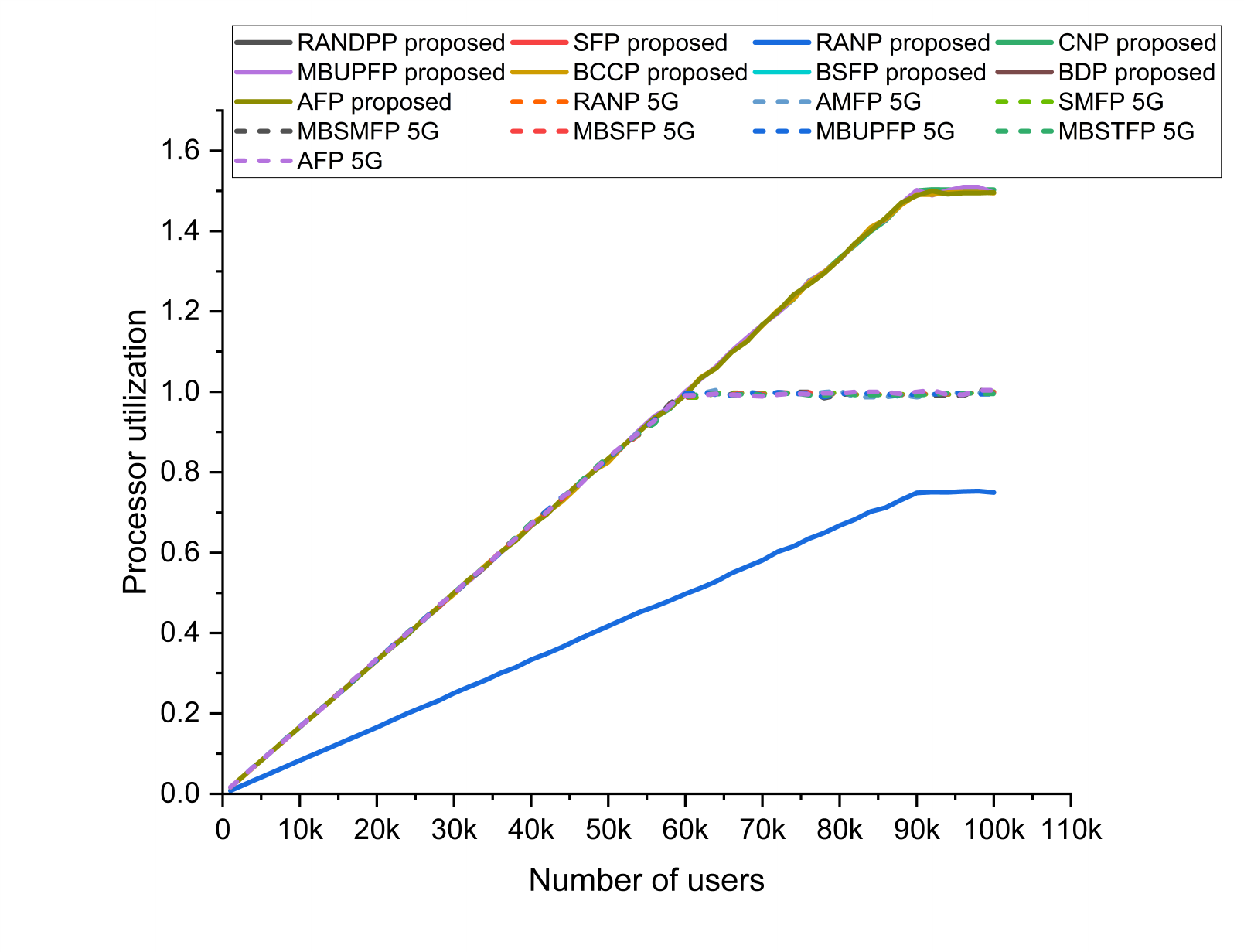}
	\vspace*{-0.3cm}
	\caption{Processor utilization of MBS session establishment for the scaled configuration.}
	\label{serv_util_3}
	\vspace{-0.2cm}
\end{figure}

\begin{figure}[h!]
	\centering
    \vspace{-0.3cm}
	\includegraphics[width=0.95\columnwidth]{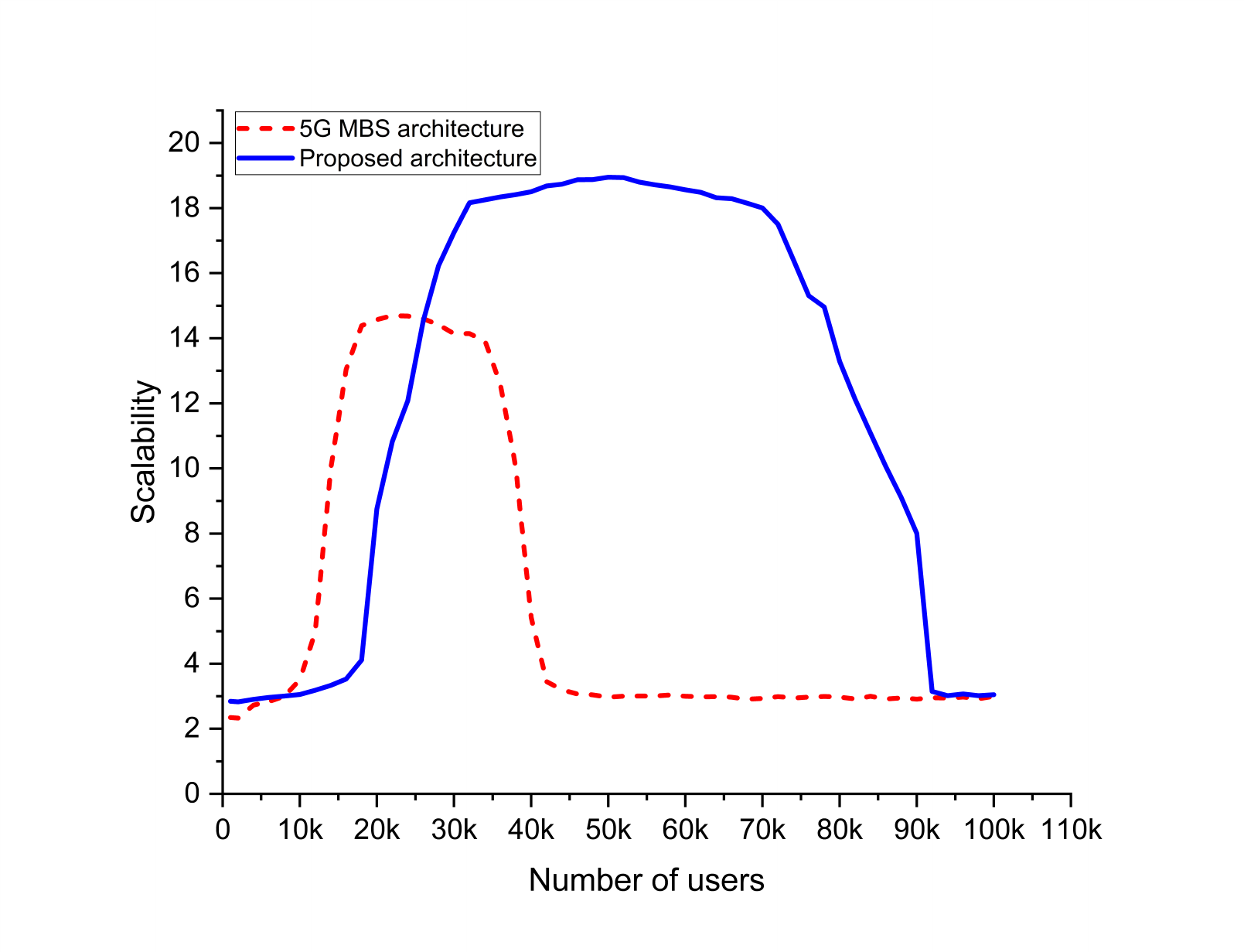}
	\vspace*{-0.3cm}
	\caption{MBS session establishment scalability.}
	\label{serv_sca}
	\vspace{-0.3cm}
\end{figure}

The comparative analysis of processor utilization of both the 3GPP 5G and the proposed SSBA for the basic configuration is shown in Fig.\,\ref{serv_util_1}. 
It shows that the BCCP (BCC Processor), for instance, reaches its maximum processor utilization, explaining the saturation point for the number of MBS session establishments. Other NFs, however, are not fully utilized at this point. This signifies that the processing chain fails if an NF becomes a bottleneck in the consecutive chain. Fig.\,\ref{serv_util_3} presents the processor utilization results for both architectures for the scaled configuration. The results demonstrate that processors in the 3GPP 5G architecture saturate earlier than the proposed SSBA due to the higher number of messages in the 3GPP 5G architecture.

Based on the obtained results for the MBS session rate, ART, and processor utilization, the scalability is evaluated using the equation provided in \cite{b22}. The scalability results are plotted in Fig.\,\ref{serv_sca} for configurations $b_1$ and $b_2$. The proposed SSBA outperforms the 3GPP 5G architecture, as it can serve more concurrent users with the same scaling configuration. 
It is evident from the results that the proposed SSBA is more scalable and performs better than the 3GPP 5G architecture.

\section{Convergence of 3GPP 5G and N3BN in SSBA}
\label{dtt}
\par
\begin{figure}[htbp]
    \hspace{0.2cm}
    \centering
    \includegraphics[scale=0.6]{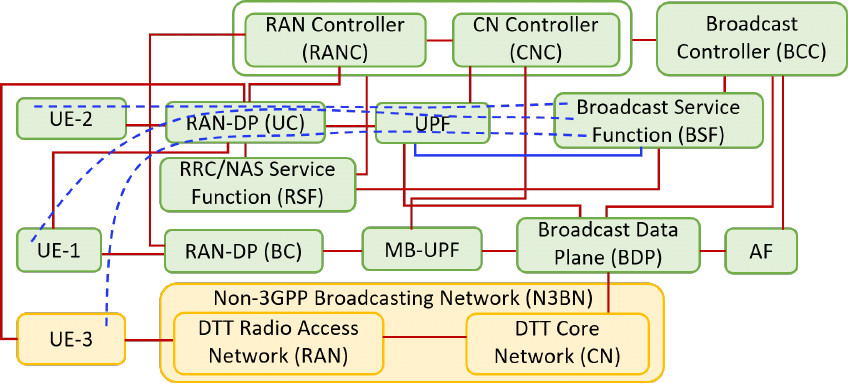}
    \caption{Convergence of N3BN (DTT) in the proposed SSBA.}
    \label{dtt_pro}
    \vspace{-0.3cm}
    \end{figure}

 The proposed SSBA can also be extended to converge 3GPP 5G cellular broadband and N3BN (say, DTT) easily. The simplified integration of N3BN (DTT) with 3GPP 5G is shown in Fig.\,\ref{dtt_pro}. The MBS-related signalling is taken care of by the BSF. However, the data path in the case of DTT content delivery is as follows: AF - BDP - DTT CN - DTT RAN - UE-3. The BSF serves as the integration point for the signalling interplay between 3GPP 5G and the DTT broadcasting network, while the BDP emerges as the point for data path integration. 

\section{Conclusion}
\label{conc}
In this paper, we have proposed a signalling service-based architecture for MBS that offers enhanced flexibility to select a delivery method based on resource availability, and results in improved network scalability by handling UE signalling as a service. Besides, it also facilitates the convergence of 3GPP 5G cellular broadband and N3BNs in the landscape of B5G. The simulations and performance evaluations were performed to demonstrate that the proposed SSBA outperforms 3GPP 5G architecture, exhibiting enhanced modularity, scalability, and a reduced number of signalling messages in the MBS session establishment procedure. In the future, we would like to perform the evaluation of the converged SSBA mobile network. 

\section*{Acknowledgment}
We acknowledge the Ministry of Electronics and Information Technology (MeitY), Govt. of India for supporting the project.

\bibstyle{IEEEtran}
\bibliography{MBS}

\end{document}